\documentstyle[11pt]{article}
\begin{document}
\title{Bispinor Formulation of Spin 3/2 Field Theory}
\author{Haryanto M. Siahaan\footnote{E-mail : hm\_siahaan@gmail.com}\\}
\date{}
\maketitle

\begin{abstract}
In this paper, we investigate an alternative formulation for spin 3/2 field equation. First we will review equation of motion of Dirac and Maxwell, and then construct the equation for spin 3/2 in the similar fashion.Our method actually a generalization of relativistic equation of motion based on spin operator and Hamiltonian similar to Dirac and Maxwell equation in ref. [1] pp. 50. Furthermore, we could not bring the equation to the Klein Gordon limit for every wave function component in spin 3/2 formalism. 
\end{abstract}

\section{Introduction}
The famous Dirac equation can be read as 
\begin{equation}
	\left(i\gamma^{\mu}\partial_{\mu}-\beta m\right)=0
\end{equation}
which also can be written as 
\begin{equation}
	i\frac{\partial}{\partial t}\psi=\left(\frac{1}{i}\vec{\alpha}\cdot\nabla-\beta m\right)\psi
\end{equation}
Then$ \left(\frac{1}{i}\vec{\alpha}\cdot\nabla-\beta m\right)$ is recognized as Dirac Hamiltonian. Each matrix in equation (2) is 
\begin{equation}
	\vec{\alpha}=\left(\matrix{&0&\vec{\sigma}\cr&\vec{\sigma}&0}\right)  ;  \beta=\left(\matrix{&I&0\cr&0&I}\right)
\end{equation}
where $\vec{\sigma}$ are Pauli matrices. In bispinor context, equation (2) can be written as a coupled equation which relate the particle wave function and it's antiparticle.
\begin{eqnarray}
	i\frac{\partial}{\partial t}\varphi=m\varphi+\frac{2}{i}\vec{S}\cdot\nabla\chi
\end{eqnarray}
\begin{eqnarray}
	i\frac{\partial}{\partial t}\chi=-m\chi+\frac{2}{i}\vec{S}\cdot\nabla\varphi
\end{eqnarray}
 where $\vec{S}=\frac{\vec{\sigma}}{2}$ as our spin operator for every two component spinor in DIrac equation. This spin operator satisfy Lie algebra for $SU(2)$ group, $\left[S^{i},S^{j}\right]=i\epsilon^{ijk}S^{k}$. The square of this operator also satisfy $\vec{S}\cdot\vec{S}=s(s+1)=\frac{3}{4}$ for $s=\frac{1}{2}$.
\par
To get the plane wave with $E^{2}=\vec{p}^{2}+m^{2}$ as a solution, equation (2) should satisfy the Klein Gordon equation. This can be done by square each operator at left hand side and right hand side. The properties of $\left\{\vec{\alpha},\beta\right\}=0$ and $\left\{\alpha_{i},\alpha_{j}\right\}=0$ which guarantee that we can get the Klein Gordon expression for each $\psi$.
\par
A similar method can be applied for Maxwell equation. In the case of no source, Maxwell equation can be read as
\begin{equation}
	\nabla\times\vec{E}+\frac{\partial \vec{B}}{\partial t}=0,
\end{equation}
\begin{equation}
	\nabla\times\vec{B}-\frac{\partial \vec{E}}{\partial t}=0.
\end{equation}
Now, by introducing such a spinor for electromagnetic fields that is $\psi_{EM}=\left(\vec{E},i\vec{B}\right)$, so equation (6)and (7) can be rewritten as 
\begin{equation}
	i\frac{\partial}{\partial t}\psi_{EM}=\frac{1}{i}\vec{\alpha}_{EM}\cdot\nabla\psi_{EM}
\end{equation}
where
\begin{equation}
	\vec{\alpha}_{EM}=\left(
\matrix{&0&\vec{S}_{EM}\cr&\vec{S}_{EM}&0}
\right)
\end{equation}
and $\vec{S}_{EM}$ is generator of rotation in 3 dimensional Euclidean space. In this way, we can say that $\left(\frac{1}{i}\vec{\alpha}_{EM}\cdot\nabla\right)$ is the "Maxwell Hamiltonian". The $\vec{S}_{EM}$can be recognized as for spin operator for $s=1$, because it satisfy $\left[S^{i},S^{j}\right]=i\epsilon^{ijk}S^{k}$ and $\vec{S}\cdot\vec{S}=s\left(s+1\right)=2$.
\begin{equation}
	S^{1}_{EM}=\left(\matrix{&0&0&0\cr&0&0&-i\cr&0&i&0}\right), S^{2}_{EM}=\left(\matrix{&0&0&i\cr&0&0&0\cr&-i&0&0}\right), S^{3}_{EM}=\left(\matrix{&0&-i&0\cr&i&0&-i\cr&0&0&0}\right)
\end{equation}
\par
Furthermore, we can check that equation (8) does not satisfy the Klein Gordon equation. In other words, each component of $\psi_{EM}$ does not have plane wave solution with $E^{2}=\vec{p}^{2}+m^{2}$ as the eigenvalue of energy. It is clear because at elemntary level we know that the dynamical variable is $A^{\mu}\left(x\right)=\left(A^{0}\left(x\right),\vec{A}\left(x\right)\right)$ which has relation with $\vec{E}\left(x\right)$ and $\vec{B}\left(x\right)$ by $\vec{E}\left(x\right)=-\left(\frac{\partial{\vec{A}\left(x\right)}}{\partial t}-\nabla A^{0}\left(x\right)\right)$ and $\vec{B}\left(x\right)=\nabla\times\vec{A}\left(x\right)$. By imposing this relation and the Lorentz gauge constraint\footnote{This constraint must be included}, we can get the massless Klein Gordon expression for every $A^{\mu}$'s component. At this point, we can learn that the Klein Gordon equation, or in other words relativistic mass-energy-momentum relation, could not be directly satisfied when we construct a higher spin formulation in "bispinor" fashion.

\section{Construction of Spin 3/2 Field Equation}
Learning from the pattern that we get in Dirac and Maxwell formalism, we can make a generalization for arbitrary spin by denoting that Hamiltonian operator could have the form $\left(\frac{a}{i}\vec{\alpha}_{s}\cdot\nabla-bm\right)$ with undetermined complex matrices a and b. $\vec{\alpha}_{s}$ is a left diagonal matrix which contains spin operator and s denote the value of spin. So in general we could have an equation which can be read as
\begin{equation}
	i\frac{\partial}{\partial t}\Psi_{s}\left(x\right)=\left(\frac{a}{i}\vec{\alpha}_{s}\cdot\nabla+bm\right)\Psi_{s}\left(x\right)
\end{equation}
where $\Psi_{s}\left(x\right)$ is the bispinor for arbitrary spin field and has $2s\left(s+1\right)$ components.
\par
So in spin 3/2 formalism, we would get an equation of motion that written as 
\begin{equation}
	i\frac{\partial}{\partial t}\Psi_{3/2}\left(x\right)=\left(\frac{a}{i}\vec{\alpha}_{3/2}\cdot\nabla+bm\right)\Psi_{3/2}\left(x\right)
\end{equation}
where
\begin{equation}
	\vec{\alpha}_{3/2}=\left(\matrix{&0&\vec{S}_{3/2}\cr&\vec{S}_{3/2}&0}\right)
\end{equation}
and $\vec{S}_{3/2}=\left(S^{1}_{3/2},S^{2}_{3/2},S^{3}_{3/2}\right)$.
\begin{eqnarray}
	S^{1}_{3/2}=\frac{1}{2}\left(\matrix{&0&\sqrt{3}&0&0\cr&\sqrt{3}&0&2&0\cr&0&2&0&\sqrt{3}\cr&0&0&\sqrt{3}&0}\right) ,\\
	S^{2}_{3/2}=\frac{i}{2}\left(\matrix{&0&-\sqrt{3}&0&0\cr&\sqrt{3}&0&-2&0\cr&0&2&0&-\sqrt{3}\cr&0&0&\sqrt{3}&0}\right) ,\\
	S^{3}_{3/2}=\left(\matrix{&3&0&0&0\cr&0&1&0&0\cr&0&0&-1&0\cr&0&0&0&-3}\right).
\end{eqnarray}
Matrices of spin operator are produced by usual angular momentum relations,
\begin{eqnarray}
	S_{\pm}=S^{1}\pm iS^{2},\\
	S^{3}\left|s,m\right>=m\left|s,m\right>,\\
	S_{\pm}\left|s,m\right>=\sqrt{\left(s\mp\ m\right)\left(s\pm\ m+1\right)}\left|s,m\pm1\right>.
\end{eqnarray}
\par
For simplification, we try to evaluate equation (12) in massless case. In an explicit form, we can write our equation as 
\begin{equation}
	i\frac{\partial}{\partial t}\left(\matrix{&\Phi\left(x\right)\cr&\Omega\left(x\right)}\right)=\frac{a}{i}\left(\matrix{&0&\vec{S}_{3/2}\cdot\nabla\cr&\vec{S}_{3/2}\cdot\nabla&0}\right)\left(\matrix{&\Phi\left(x\right)\cr&\Omega\left(x\right)}\right).
\end{equation}
Before further investigation, we should recognize that our spin operator, or more precisely our $\vec{\alpha}_{3/2}$ do not anticommut for each different components\footnote{Compared to Dirac's $\vec{\alpha}$} and $\alpha^{i}_{3/2}\alpha^{i}_{3/2}\neq\frac{1}{a}I_{4\times4}$ which means equation (20) cannot simply has Klein Gordon expression. Our formalism cannot directly fulfill this requirement to be such a relativistic equation. Probably this problem is similar to Maxwell formalism when we use the "spinor" $\psi_{EM}=\left(\vec{E},i\vec{B}\right)$ as the dynamical variable. Perhaps we need another constraints to satisfy this relativistic wave equation requirement or finding another another"\textit{dynamical variable}" that can be related to our bispinor for describing spin 3/2 fields just like in Maxwell formalism case before.

\section{Conclusion}
Trying to get such a consistent theory of higher spin fields (higher than spin 1) is a complicated manner. Beside, there still no single particle with higher spin that needs some explanation ever exist in laboratory. But effort to get a good equation is an interesting thing. Furthermore, finding the expression (11) for standard Rarita Schwinger formulation can be proved imposible.
\begin{center}
***
\end{center}
\par
I am grateful to Prof. Triyanta for interesting discussion and his encouragement.

\section{References}
\begin{enumerate}
	\item C. Itzykson, and J. Zuber, "Quantum Field Theory,"McGraw-Hill (1980). pp. 48-50.
	\item W.Rarita, and J. Schwinger, Phys. Rev. 60 (1941) 61.
\end{enumerate}

\end{document}